\documentstyle[12pt]{article}
\def\be{\nopagebreak[3]\begin{equation}}
\def\ee{\end{equation}}
\def\ba{\nopagebreak[3]\begin{eqnarray}}
\def\ea{\end{eqnarray}}
\def\nl{\nonumber \\}
\def\ni{\noindent}

\begin{document}
\begin{titlepage}
\pagestyle{empty}

\title{
A HAMILTONIAN FORMULATION OF TOPOLOGICAL
 GRAVITY\thanks{This research was supported in part by
the National Science Foundation Grant No. PHY89-04035, by CONACyT (Mexico)
and by the A.G.C.D. (Belgium).}
}
\author{
 H. WAELBROECK\thanks{Permanent address: Inst. C. Nucleares,
UNAM, Circ Ext, C.U., Mexico D.F. 04510}
and J. A. ZAPATA\thanks{P. a.: Dept. of Physics,
Penn. St. University, 104 Davey Lab, Univ. Park, PA
16802}\\Institute for Theoretical Physics,\\University of
California at Santa Barbara,\\Santa Barbara, CA.
}

\maketitle

\vspace{1cm}
\begin{abstract}

Topological gravity is the reduction of Einstein's theory to spacetimes with
vanishing curvature, but with global degrees of freedom related
to the topology of the universe. We present an exact Hamiltonian lattice theory
for topological gravity, which admits translations
of the lattice sites as a gauge symmetry. There are additional
symmetries, not present in Einstein's theory, which kill the local
degrees of freedom. We show that these symmetries can be fixed by choosing
a gauge where the torsion is equal to zero. In this gauge, the theory
describes flat space-times. We propose two
methods to advance towards
the holy grail of lattice gravity: A Hamiltonian lattice theory for curved
space-times, with first-class translation constraints.

\end{abstract}
\vfill
\end{titlepage}

\section{Introduction}
The nonlinearity of Einstein's equations imply that, in many situations
of physical interest, one must turn to nonperturbative methods,
typically numerical computations based on a lattice theory \cite{1}. For
similar reasons, perturbative quantum gravity is plagued with
nonrenormalizable infinities \cite{2}, a fact which has convinced many that
a nonperturbative  approach \cite{3}
must be developped before one can even discuss
the existence of the theory. In view of the difficulties associated
with nonperturbative quantum field theory, physicists have turned
increasingly to alternative discrete
theories \cite{4}, among them Hamiltonian
lattice theories \cite{5}.

In both, numerical relativity, and lattice theories of quantum gravity,
a key problem has been that of finding constraints, analogous to the
diffeomorphism constraints ${\cal H}_{\mu}(x) \approx 0$, that are consistent
with time evolution even for a finite lattice. In the constrained Hamiltonian
formalism, this requires that the constraints form a closed , or
``first-class'', algebra \cite{6}. Indeed,

\ba
\lbrack H , {\cal H}_{\nu} \rbrack &=& \lbrack {\sum _x}N^{\mu}(x) {\cal H}
_{\mu}(x) , {\cal H}_{\nu} \rbrack \nl
&\approx & \sum N^{\mu}(x) \lbrack {\cal H}_{\mu}(x) , {\cal H}_{\nu}
\rbrack \label{1}
\ea
\ni
which vanishes for all $\{x,N^{\mu}(x)\}$ if and only if  $\lbrack {\cal
H}_{\mu} , {\cal H}
_{\nu} \rbrack \approx 0 $.

One can achieve consistency with time evolution for {\em fixed $N^{\mu}(x)$} by
solving (\ref{1}) for these lapse-shift functions. However, this would increase
the number of
degrees of freedom per point, since the number of symmetries would be reduced.
Also when the lattice brackets analogous to
$\lbrack {\cal H}_{\mu} , {\cal H}
_{\nu} \rbrack $ are non-zero, they are usually very small, so that
${\cal H}_{\mu}$ generates a flow which is almost a symmetry of the action,
which will be nearly flat along this flow. This causes a problem both for the
convergence of numerical relativity schemes, and for the calculation of the
quantum gravity path integral. In light of these facts, the task of finding
a Hamiltonian lattice theory with first-class constraints has drawn a
great deal of attention in the past decades \cite{5}.
Believing that the problem
is too difficult to tackle head on, we have turned our attention first to
the model of
2+1-dimensional gravity \cite{7}, \cite{8}, since in this model
the space-time curvature
vanishes \cite{9}, and the quantum theory can be formulated elegantly
\cite{10}. We then showed how curvature could be introduced through
the reality conditions in the complexified theory \cite{11}.
We are now aiming to
generalize this work to 3+1-dimensional space-times.

The purpose of this article is to present a lattice theory for topological
gravity, with first-class constraints. To achieve this, we generalize
previous work in 2+1 dimensions by exploiting Horowitz's
results, which show that a generalization
of Witten's work from $2+1$ to $3+1$ dimensions \cite{12}, leads to
a topological theory for flat spacetimes coupled to closed two-forms.

Horowitz proposed going from the Palatini action for Einstein gravity
\cite{13},

\be
S = \int _{\cal M} e^a \wedge e^b \wedge R^{cd} \varepsilon _{abcd} \quad ,
\label{2}
\ee

\ni
to the topological action, replacing $(e\wedge e)^*$ by a two form $B$.

\be
S_T = \int _{\cal M} B_{cd} \wedge R^{cd} \quad .
\label{3}
\ee

\ni
One can consider (\ref{3}) as an action functional over the fields $B$ and
an $SO(3,1)$ connection
$\omega$ $(R=d\omega +[\omega ,\omega ])$, and drop the constraint

\be
B^* = e \wedge e \quad .
\label{4}
\ee

The action (\ref{3}) should give a larger set of solutions
than (\ref{2}), since there are fewer constraints. Yet
at first sight (\ref{3}) has only global degrees of freedom. Indeed,
the variational
equations from (\ref{3}) state that $\omega$ is a flat $SO(3,1)$ connection,
and $B$ a closed two-form with values in $SO(3,1)$. Furthermore, (\ref{3})
is invariant under the usual Maxwell
gauge transformations and under the translations of $B$
by an exact form. Thus, the space of solutions is the set of inequivalent
flat $SO(3,1)$ connections on $\cal M$, together with the inequivalent
closed two-forms. These are related to the first and second cohomology
vector spaces $H_1({\cal M})$ and $H_2({\cal M})$. For ${\cal M}=\Sigma
\times (0,1)$ the existence of a canonical structure rests on the Poincar\'e
duality between the vector spaces $H_1(\Sigma)$ and $H_2(\Sigma)$. The
local degrees
of freedom of (\ref{2}), that do not appear in the topological action
(\ref{3}), are hidden in the torsion, as Horowitz  pointed out. From this point
of view, topological gravity is seen to be a
theory of teleparallelism \cite{14}, where one chooses a
connection with torsion
but no curvature, rather than the contrary%
\nopagebreak[1]\footnote{ Given any pseudoriemannian metric one can construct a
{\em flat} connection
compatible with the metric; but it is usually a  connection
with torsion. This implies that every space time is a solution
of (\ref{3}); the converse is not necessarily true \cite{15}}.

The lattice theory, analogous to Horowitz's topological gravity,
is based on the following
lattice analogues of $\omega $ and $B$:
A set of $SO(3,1)$ matrices, which define parallel-transport through each
lattice face, and the area bivector of each face.
The first-class constraints demand the closure of each lattice
cell (the sum of area bivectors on the boundary of a cell
must be zero) and the vanishing of the
curvature at each bone of the lattice. The former generate $SO(3,1)$
transformations of the frame at
each cell, while the latter split into two groups: The generators of
spacetime translations at each lattice site, which reduce to ${\cal H}_{\mu}
(x)$ in the continuum limit, and generators of additional symmetries
which are responsible for eliminating the local degrees of freedom of
Einstein's theory. We will fix these last symmetries by demanding
that the
lattice analogue of torsion vanish, thereby reducing the theory to
flat spacetimes.

The article is organized as follows. The lattice theory is introduced in
section two, by constructing a lattice analogue of the action (\ref{3}). In
section three, we will impose ``geometricity conditions,'' which imply that
the variables represent a flat spacetime (with no torsion).
The possibility of extending this work to space times with
curvature (or torsion) is discussed in the concluding section.

\def\bomega{\mbox{\boldmath $\omega$}}

\section{Topological Lattice Theory}
For topological theories, which have only a finite number
of degrees of freedom, it is logical to put the
settings of the theory in a discrete context. We follow this approach, and
formulate a theory for topological gravity that in a certain gauge
has the interpretation of a geometrical lattice.

One can think of the variables of the discrete $3+1$-dimensional gravity
as the result of integrating the continuum ones in a space-like
direction. That is, instead of the connection $\omega$ we take Lorentz
matrices ${\bf M}$ which arise from integrating the connection in a space-like
direction. Such matrices
realize parallel transport along noninfinitesimal distances.
In the place of the bivector valued two-form $B$ we consider its integral
over noninfinitesimal areas, and denote these by ${\bf E}$. The continuous
variable $\omega _0$
has no discrete counterpart because it describes a time-like aspect of the
continuous theory and we are seeking a discrete theory with continuous time.
For the same reason the variable $B_{k0}$
has a discrete-continuous counterpart $B(I)^{[\mu \nu ]}=
(\int _{l(I)}e_i^{[\mu }dx^i)e_0^{\nu ]} = A(I)^{[\mu } e_0^{\nu ]}$,
were $A(I)$ is a spacelike noninfinitesimal segment. There are two important
continuous quantities which have discrete counterparts that may also be
derived from the discrete variables. The discrete counterpart of the curvature
two-form is its integral over non-infinitesimal surfaces dual to the segment
$I$: $P(I)$. In the
place of the exterior derivative of the two-form $B$ we take its integral
over the boundary of a boundary of a non-infinitesimal volume $i$: $J(i)$.

If one separates the time and spatial directions and integrates by parts one
of the resulting terms, the continuum action for topological gravity (\ref{})
can be rewritten in a form which makes it easier to change to the
Hamiltonian formalism and identify the discrete
counterparts of each term.

\be
S_T= 3 \int dt \int _{\Sigma }(\dot \omega ^A_{[i} B_{jk]A}-
R^A_{[ij}B_{k]0A} +\omega ^A_{0} D_{[i} B_{jk]A})dx^idx^jdx^k
\label{}
\ee

\ni

The Lagrangian has a natural discrete analog which will be our starting point

\be
L={\sum _{ij}}{\bf E}_{ij}\cdot \dot {\bomega}_{ij} -
{\sum _{I}}{\bf P}(I)\cdot B(I) +
{\sum _{i}}\omega (i)_{0}\cdot J(i)
\label{6}
\ee

\ni
With the help of a little symbolic manipulation it is possible to
rearrange the first term%
\nopagebreak[3]\footnote{Equation (\ref{8}) is strictly correct only for an
Abelian
group, since it neglects the ordering ambiguity of the two matrices.
However, it is easy to see that there are only two ways to write the
dynamical term (\ref{L}) considering that indices can be contracted only
if they live in the same frame, the only other possibility being equivalent
to this after integration by parts.}
:

\ba
{\bomega} &=& ln {\bf M}\\
\dot {\bomega} &=& {\bf M}^{-1}\dot {\bf M} \label{8} \\
L&=&{\sum _{ij}} C^B_{\ C A} E_{(ij)}^{\ \ \ A}M_{(ji)\ \ B}
^{\ \ \ D},\dot M_{(ij)\ \ D}^{\ \ \ C}
 +...
\label{L}
\ea

\ni
where $C^B_{\ C A}$ are the structure constants of the Lie algebra of
$SO(3,1)$; they are also the generators of the Lorentz transformations
for bivectors, i.e. $M_{(ij)\ \ B}^{\ \ \ A}= exp(\omega _{(ij)\ C}
C^{CA}_{\ \ B})$. The latin indices $(i,j,\ldots )$ denote lattice cells,
and a pair of indices $(ij)$ denotes the lattice face which separates cells
$(i)$ and $(j)$.

At this stage we consider the $12 N_2$ numbers $E_{(ij)}^{\ \ \ [ab]},
E_{(ji)}^{\ \ \ [ab]}$, in addition to the $72 N_2$ numbers
$M_{(ij)\ \ \ [cd]}^{\ \ \ [ab]},M_{(ji)\ \ \ [cd]}^{\ \ \ [ab]}$,
as independent dynamical variables.
 The first term of the Lagrangian (\ref{6}) leads to constraints linear in the
momenta, which can be solved in conjunction with the following
constraints, that restrict the matrices ${\bf M}_{ij}$ to be orthogonal
and all the variables to be antisymmetric in $(ij)$,

\ba
&&E_{(ij)}^{\ \ \ A}=- M_{(ij)\ B}^{\ \ \ A} E_{(ji)}^{\ \ \ B} \quad ,
\label{orto}\\
&&M_{(ij)\ C}^{\ \ \ A}  M_{(ji)\ B}^{\ \ \ C}= \delta ^A_{\ B} \quad ,
\label{inv}\\
&&M_{(ij)\ C}^{\ \ \ A} M_{(ij)}^{\ \ \ BC} = \eta ^{AB} \quad .
\label{anti}
\ea

\ni
The Dirac procedure gives
rise to the brackets \cite{7}

\ba
\lbrack E_{(ij)}^{\ \ \ A} , E_{(ij)}^{\ \ \ B} \rbrack &=& C^{AB}_
{\ \ D}  E_{(ij)}^{\ \ \ D} \quad ,
\label{4}
\\
\lbrack E_{(ij)}^{\ \ \ A} , M_{(ij)\ C}^{\ \ \ B} \rbrack &=& C^{AB}_
{\ \ D}  M_{(ij)\ C}^{\ \ \ D} \quad ,
\label{5}
\\
\lbrack E_{(ij)}^{\ \ \ A} , M_{(ji)\ C}^{\ \ \ B} \rbrack &=&- C
^{AD}_{\ \ \ C}  M_{(ji)\ D}^{\ \ \ B} \quad ,
\label{6}
\ea

\ni
where $C^{[ab][cd][ef]}=\varepsilon ^{[ab]r}_{\ \ \ \ \ s}
\varepsilon ^{[cd]s}_{\ \ \ \ \ t}\varepsilon ^{[ef]t}_{\ \ \ \ \ r}$
are the structure constants of the Lorentz group.

The second and third terms of the Lagrangian (\ref{6}) are
responsible of the constraints

\ba
&&J(i)^{\ \ A} =E_{(ij)}^{\ \ \ A}+E_{(ik)}^{\ \ \ A}+\ldots \approx 0 \quad ,
\label{cerra}
\\&&
P(I)^A ={1\over 4} C_B^{\ CA}
W(I)^B_{\ C}={1\over 4} C_B^{\ CA}({\bf M}_{ij}{\bf M}_{jk}\ldots {\bf
M}_{ni})^B_{\ C} \approx 0 \quad ,
\label{plani}
\ea

\ni
It is useful to notice that for matrices $W(I)$ near the the identity
we have $W(I)^A_{\ B}= exp(
C^A_{\ BC}P_{(I)}^C)$.  As we will see, within the geometrical
gauge, the constraints (\ref{cerra},\ref{plani}) can be interpreted as
the requirement that the faces close and that the curvature of the connection
vanishes.
For these brackets the restrictions (\ref{orto})-(\ref{anti}) are identities,
and one can show that the constraints (\ref{cerra}) and (\ref{plani})
are first-class. The first
generates Lorentz transformations, and the second,
translations of the bivectors ${\bf E}$
by the analogous of an exact form, which in the lattice context means a
translation of the bivectors in a way that does not break the closure
conditions. In the next section,
we will show how some of these transformations are related to translations
of lattice sites.

\ba
\lbrack J_{(i)}^{\ \ A} , E_{(ij)}^{\ \ \ B} \rbrack &=& C^{AB}_
{\ \ D}  E_{(ij)}^{\ \ \ D} \quad ,
\label{lor1}\\
\lbrack J_{(i)}^{\ \ A} , M_{(ij)\ C}^{\ \ \ B} \rbrack &=& C^{AB}_
{\ \ D}  M_{(ij)\ C}^{\ \ \ D} \quad ,
\label{lor2}\\
\lbrack \xi ^A P_{(I)\ A} , E_{(ij)}^{\ \ \ B} \rbrack &\approx &\xi ^B \quad .
\label{tra}
\ea

Not all the translations generated by $P(I)$ are independent: The Bianchi
identities imply
that the sum of the flatness conditions (\ref{plani}) associated
to all the links that
flow into a given vertex, is redundant.
We now know the number of
variables, the number of constraints and the number of symmetries they
generate. Hence, we can
now compute the total number of configuration space degrees freedom of
the theory:

\be
d=6N_2 - 6N_3 -6(N_1 - N_0)=6\chi \quad .
\label{0}
\ee

This is just six times the Euler number $(\chi )$, which is always equal to
zero
in three dimensions because of the Poincar\'e duality for the Betti numbers,
$b_i=b_{n-i}$.

\be
\chi =b_0-b_1+b_2-b_3=0
\label{22}
\ee

Thus, all of the degrees of freedom of the lattice can be gauged away.
The counting given above fails at the global level when, for some topologies,
some of the constraints become redundant. This implies that, for
certain topologies, $3+1$ dimensional flat spacetimes can have degrees
of freedom. We discuss this further in a separate article \cite{15}, which
describes a reduced version of this theory with a minimal lattice,
which has only one vertex and one cell $(N_0=N_1=1)$. The reduction of the
lattice skeleton to minimal form follows the same procedure as in $2+1$
dimensional gravity \cite{16}.

\section{Geometrical Gauge and \hspace{10cm}
\-Translation \-Symmetry}
By the geometrical gauge we mean a gauge in which the configuration variables
of the theory, ${\bf E}_{ij}$ are area bivectors for the faces of a
simplicial lattice, with vertices that are
linked by straight
lines, which can be described by four-vectors ${\bf A}(I)$.
The bivectors ${\bf E}^*_{ij}$ are then
expressed as a wedge product of the vectors $A(J)$, $A(K)$ and $A(L)$ that
correspond to the frontier of the face $(ij)$

\be
{\bf E}^*_{ij} ={\bf A}(J)\wedge {\bf A}(K)= {\bf A}(K)\wedge {\bf A}(L)=
{\bf A}(L)\wedge {\bf A}(J) \quad .
\label{E=AA}
\ee

\ni
Since the linking vectors ${\bf A}$ form the boundary of the face $(ij)$, they
satisfy the closure conditions

\be
{\bf A}(J)+{\bf A}(K)+{\bf A}(L)=0
\ee

Equation (\ref{E=AA}) leads to a first set of geometricity conditions,
that guarantee the geometricity of each separate
cell: The requirements that the bivectors
${\bf E}^*$ represent
a plane face between neighboring cells, and that the faces of a cell
intersect in pairs, so that each cell be totally contained in a
three-dimensional subspace of Minkowski space-time.
These geometricity conditions, for cell $(i)$ are

\ba
F(ij) &\equiv & \varepsilon _{abcd} E_{(ij)}^{\ \ \ ab} E_{(ij)}^{\ \ \ cd}
= 0 \quad ,\label{Fij}\\
C(ijk) &\equiv & \varepsilon _{abcd} E_{(ij)}^{\ \ \ ab} E_{(ik)}^{\ \ \ cd}
= 0 \quad .\label{Cijk}
\ea

We also want to have a covariant description in which parallel transport
between neighboring faces is described by Lorentz matrices
$M_{(ij)\ b}^{\ \ \ a}$ %
\footnote{We will use the same notation $(M_{(ij)\ b}^{\ \ \ a})$
for these matrices, which act on vectors, as for the previously defined
$(M_{(ij)\ B}^{\ \ \ A})$ in the bivector representation; it is the same
element of $SO(3,1)$, but in a different representation.}.
We refer to the variables ${\bf A}(I)$ and ${\bf M}_{ij}$ as the geometrical
variables.
The geometricity conditions for the variables of the theory must imply that
the geometrical variables satisfy the compatibility conditions

\be
{\bf A}(I_i)=-{\bf M}_{ij}{\bf A}(I_j)
\label{A=-MA}
\ee

\ni
where ${\bf A}(I_i)$ is the vector asociated to the link (I) in the frame (i),
and ${\bf A}(I_j)$ is the same vector, parallel transported to the frame $(j)$.

This geometrical requirement will be turned into constraints for the variables
${\bf E, M}$ of the theory. Some of these requirements are contained in the
identity ${\bf E}_{ij}=-{\bf M}_{ij} {\bf E}_{ji}$%
\footnote{The identities ${\bf E}_{ij}=-{\bf M}_{ij} {\bf E}_{ji}$ and
${\bf E}^*_{ij}=-{\bf M}_{ij} {\bf E}^*_{ji}$ are equivalent.}
, but there are other conditions. These new restrictions
reflect the different ways that ${\bf E}^*$ can be written as ${\bf A}\wedge
{\bf A}$;
in particular they reflect the degree of freedom corresponding to the rotations
within
the plane defined by ${\bf E}^*$. The constraint on the connection matrices
which freezes this degree of freedom
is just the requirement that the lattice torsion vanishes.
This requires that links on the boundary between two
cells be the same when seen from either cell, i.e. that if ${\bf A}(I_i)$
is the intersection of the faces ${\bf E}_{ij}^*$ and ${\bf E}_{ik}^*$ of cell
$(i)$, and ${\bf A}(I_j)$ is the intersection of two faces of cell $(j)$,
${\bf E}_{ji}^*$ and ${\bf E}_{jl}^*$, then equation (\ref{A=-MA}) holds:
${\bf A}(I_i)=-{\bf M}_{ij}{\bf A}(I_j)$. Since one has
${\bf E}_{ij}^*=
-{\bf M}_{ij}{\bf E}_{ji}^*$, it is sufficient that the planes defined by
${\bf E}_{ij}^*$, ${\bf E}_{ik}^*$ and ${\bf M}_{ij}{\bf E}_{jl}^*$ all
intersect along the same line:

\ba
C_{ABC}E_{(ij)}^{\ \ \ A} E_{(ik)}^{\ \ \ B} M_{(ij)\ D}^{\ \ \ C}
E_{(jl)}^{\ \ \ D} = 0 \quad .
\label{29}
\ea

\ni
These conditions, together with (\ref{Fij}, \ref{Cijk}), guarantee
the geometricity of the lattice. They form a redundant set of constraints,
which fix all but $4N_0$ of the flatness constraints ${\bf P}(I)\approx 0$.
The remaining first-class constraints generate translations of each lattice
vertex $(v)$ in the embedding Minkowski space, and are given by

\ba
H_a(v) = {\sum _{I\rightarrow v}} \varepsilon _{abcd} A(I)^b
P(I)^{cd}\label{T}
\ea

\ni
where ${\bf A}(I)$ is the vector associated to the link $(I)$ and pointing
to the vertex $(v)$. Within the
geometrical gauge one can obtain the vectors
${\bf A}(I)={\bf A}(ijk)$ as functions of the
bivectors ${\bf E}_{ij}^*$ and ${\bf E}_{ik}^*$, which define the direction of
${\bf A}(I)$ as theier intersection place, and as function of a 3-volume
element for cell $(i)$ which is a function of the four bivectors
${\bf E}^*$ that conform cell $(i)$.

\be
A(I)^a=-\varepsilon (i)^{abc} \varepsilon (i)_{bef} E_{(ij)}^{*\ \ ef}
\varepsilon (i)_{cgh} E_{(ik)}^{*\ \ gh} \quad ,
\label{A}
\ee
\ni
were
\ba
&& \varepsilon (i)_{abc}={\phi (i)_{abc}\over \sqrt{
{1\over 6}\phi (i)_{efg}\phi (i)^{efg}}} \quad ,\\
&& \phi (i)_{abc}= \varepsilon _{fghk}
E_{(ij)\ [a}^{*\ \ f} E_{(ik)\ b}^{*\ \ g} E_{(il)\ c]}^{*\ \ h}
x(i)^k \quad ,
\ea

\ni
where ${\bf x}(i)$ is a four-vector that is transverse to the tetrahedron
$(i)$. One can show that that the 3-volume element $\varepsilon (i)$ doesn't
depend on the choice of ${\bf x}(i)$.

Now it is possible to
get an expression for the translation generators in terms of the bivectors
${\bf E}$.
The action of the constraints $H(v)_a$ on the bivector
${\bf E}_{ij}^*={\bf A}(J)
\wedge {\bf A}(K)={\bf A}(K) \wedge {\bf A}(L)={\bf A}(L)\wedge {\bf A}(J)$,
which contains the vertex $(v)$ as the place of intersection of ${\bf A}(J)$
and ${\bf A}(K)$, is given by ($N^a$ is the ``lapse-shift''
vector for vertex $(v)$)

\ba
\lbrack N^a H_a(v) , E_{\ (ij)}^{*\ \ \ ef} \rbrack
&\approx & \varepsilon _{abcd} \varepsilon ^{ef}_{\ \ gh}
N^a A(J)^b \lbrack P(J)^{cd} , E_{(ij)}^{\ \ \ gh} \rbrack \nl
&+& \varepsilon _{abcd} \varepsilon ^{ef}_{\ \ gh}
N^a A(K)^b \lbrack P(K)^{cd} , E_{(ij)}^{\ \ \ gh} \rbrack \nl
&=& N^a ({1\over 2}
 A(J)^b \delta _{ab}^{\ \ ef} -
{1\over 2} A(K)^b \delta _{ab}^{\ \ ef})\nl
&=& {1\over 2} N^a \delta _{ab}^{\ \ ef} (A(J)^b - A(K)^b) =
{1\over 2} \delta _{ab}^{\ \ ef} N^a  A(L)^b
\label{Tr}
\ea

In the geometrical picture, a translation of the vertex $(v)$ by $N^a$ would
have the same effect. This means that $H_a(v)$ is the generator
of translations of the vertex $(v)$.

The claim that the translation
constraints $H_a(v)$ are the lattice counterpart of the
diffeomorphism constraints is not only based in the fact that the lattice
counterpart of diffeomorphisms are translations of the lattice vertices, it
is supported by an analogy between the dynamical components of the Einstein
tensor $G_{a0}$ and the translation constraints of the lattice.

\ba
R^a_{\ b\ cd} &\to &P(I)^a_{\ b}P(I)_{cd}/ |{\bf P}(I)| \\
G_{ab} &\to & \varepsilon ^c_{\ a[rs]}P(I)^{[rs]}
\varepsilon _{cb[tu]}P(I)^{[tu]}/|{\bf P}(I)| \quad .
\ea
\ni

In the right-hand sides of these expressions, we have used the fact that for
a geometrical lattice, a small spacelike area orthogonal to the link ${\bf
A}(I)$
can be given by
$P(I)^{ab}= \varepsilon ^{ab}_{\ \ cd} A(I)^cX(I)^d$, where $X(I)$ is some
timelike vector. Using this expression of ${\bf P}(I)$, and a four-vector ${\bf
t}$
orthogonal to ${\bf A}(I)$, we get

\be
G_{ab}t^b \to -{1\over 2} \varepsilon ^c_{ars} P(I)^{rs}
A(I)_c t^b X(I)_b/|{\bf P}(I)|
\ee

\ni
The analogy can be used to see that the continuum limit of the translation
constraints are the
diffeomorphism constraints.
If we use%
\footnote{The lattice theory in its present form describes spacetimes with
vanishing curvature. The requirement of an appropriate choice of the
vector ${\bf X}(I)$ for the continuum limit, hints that this vector
will probably be rigidly defined in a future
extension of this lattice theory to spacetimes with curvature.
However, it is important to notice that in the present theory we can
choose ${\bf X}(I)$ to be an arbitrary timelike vector.}
 the correct vector ${\bf X}(I)$ and pass to the continumm limit,
where we can choose a time four-vector ${\bf t}(v)$ orthogonal to all
the lattice links flowing into the lattice vertex $(v)$, we can evaluate
the counterpart of the Einstein tensor in the lattice vertex $(v)$ to get

\be
G_{a0}=G_{ab}t^b \to k(v) {\sum _{I\rightarrow v}} \varepsilon _{acrs} A(I)^c
P(I)^{rs}
\ee

We have arrived at a geometrical theory that can be described
using the bivectors ${\bf E}$, but which has as natural variables the
$N_1$ vectors ${\bf A}(I)$ that describe the lattice links. These vectors
satisfy a closure condition for each face, minus a redundancy for each cell.
This, in addition to the $4N_0$
symmetries of the theory, means that the total number of degrees of freedom
of the theory is%
\footnote{This counting of the degrees of freedom does not include an analysis
of the redundancies of the constraints and symmetries. This is the reason
why the degrees of freedom of some spacetimes with specific topologies do
not appear in the counting. For "generic" topologies (specifically,
those that do not admit a Seifert bundle structure \cite{15}), the
counting (\ref{36}) is correct and is related to the ``rigidity theorem'' of
Mostow \cite{104}.}

\be
4N_1 - 4 (N_2 - N_3) -4N_0 =0
\label{36}
\ee

In the original, non geometrical theory, there were other ``translational''
gauge freedoms, which did not commute with the geometricity conditions.
These were fixed by imposing the $torsion =0$ constraints. Since the $4N_0$
translation constraints commute with the $torsion =0$ conditions, one simply
chose an initial surface which is geometrical, and evolve it in time with the
hamiltonian constraints $H_a(v)$, using the original Poisson Brackets
(\ref{4}-\ref{6}).

\section{Conclusion}
The principal aim of this article has been to show how the exact translation
symmetry of 2+1 dimensional
lattice gravity carries over to 3+1 dimensions. The latticized
Chern-Simons theory in 2+1 dimensions indeed generalizes nicely to a lattice
theory for topological gravity. The latter is exactly solvable as a classical
theory, and a Hilbert space of quantum states can be identified more or less
explicitly \cite{101}, \cite{102}.

One way to understand the reduction of gravity to topological gravity, is
by taking note of the increase in the number of constraints, from four
translation constraints per point (i.e., $4N_0$) to $6(N_1-N_0)$ flatness
conditions, where $N_1$ is the number of lattice links and $N_0$ is the
number of lattice sites. As we saw, the translation constraints, analogous to
${\cal H}_{\mu} \approx 0$, can be reconstructed by forming $4N_0$ projections
from the larger set of $6(N_1-N_0)$ constraints. The remaining $6N_1-10N_0$
constraints of topological lattice gravity, are the responsible for killing
the local degrees of freedom of the theory.

This insight suggests a novel approach to the longstanding problem of finding
a lattice theory for
ordinary gravity, with $4N_0$ first-class constraints which would
reduce to ${\cal H}_{\mu}(x)$ in the continuum limit. The proposal is
summarized in the diagram below [Figure 4.1]: One first reduces to
topological gravity by increasing the number of constraints. This reduced
theory can be placed on a lattice, without breaking the translation symmetry,
as we have shown in this article. Given the lattice theory, and $4N_0$
translation constraints, one then drops the extra $6N_1-10N_0$ constraints,
which were responsible for reducing the theory to flat spacetimes.

Unfortunately, one easily shows that that the $4N_0$ translation constraints
form a closed (first-class) algebra only when the other $6N_1-10N_0$
constraints are satisfied (i.e., when the spacetime is flat). If the flatness
conditions are to be dropped, one would have to
modify the $4N_0$ constraints by
adding terms proportional to the curvature, so that the algebra of the
constraints closes even for non-vanishing curvature. Note that such extended
constraints would automatically reduce to the $4N_0$ constraints given above
when the curvature does vanish.

Finding such extended constraints is a difficult task: All we have done is
to translate the difficulty of finding first-class lattice constraints, into
this new context. However, there are general methods designed for the problem
of finding first-class extensions of constraints which close only ``on-shell''
(i.e.,
when other constraints, the ``shell-conditions,'' hold), such as that of
adding terms linear in BRST ``ghost'' variables, and the existence of a
solution to this general problem has been proved \cite{103}. This existence
proof is in itself an important point, since it implies that our failure to
find first-class constraints for lattice gravity is really our own failure,
rather than a signal that lattice gravity could not possibly admit such
symmetry. This contradicts previous arguments by various authors, including
one of us (HW), that lattice gravity should not have translation symmetry
because, in a piecewise-flat lattice, displacing
lattice sites clearly leads to a change in the metric properties of the
manifold. It is not a priori impossible that such a deformation, coupled to
an appropriate change in the parallel transport matrices, could leave a
lattice-gravity action invariant. To see just what combinations of lattice
deformations and parallel-transport changes are ``symmetries,'' is precisely
what one would accomplish by finding first-class constraints .   We stress,
however, that the method of finding first-class extensions of second-class
constraints has not been previously applied to lattice theories, as far as
we are aware; there is no guarantee that the method is practically applicable
in this context.

Another approach to the problem of first-class lattice constraints is
suggested by the theory of teleparallelism. One can view the lattice theory
without geometricity conditions, as a framework for a lattice theory
of curved spacetimes, but described with a ``connection'' that has torsion
and no curvature. From this point of view, the lack of local degrees of
freedom of topological gravity
reflects only our failure to count the torsion degrees of freedom,
where the dynamics lies. One could attempt to include lattice ``torsion''
variables and extend the constraints and bracket algebra to act non-trivially
on the new variables. The task is, again, to find first-class extensions
of the translation constraints.

Thus, the dream of finding a consistent Hamiltonian lattice gravity theory
remains very much alive; we hope that this article will turn out to be a
constructive step in that direction.

\newpage

\vfill

\section*{Figure captions}

Figure 4.1 \ \ The detour through topological gravity is suggested as a way
to find a consistent Hamiltonian lattice theory of gravity. The first
step, to define the
topological gravity, was carried out by Horowitz \cite{12}.
The second, to topological lattice gravity, was completed in this article.
The remaining task (dotted arrow) is to seek first-class extensions of
the $4N_0$ translation constraints, possibly by adding terms linear in
the curvature and the BRST ghosts.

\end{document}